\relax
\documentclass{article}
\usepackage{ijcai19}
\usepackage{times}  
\usepackage{helvet}  
\usepackage{courier}  
\usepackage{url}  
\usepackage{graphicx}  
\usepackage{footnote}
\usepackage{subfigure}
\usepackage{amsmath}
\usepackage{amsfonts}
\usepackage{multirow}
\usepackage{bm}
\usepackage{enumitem}
\usepackage{ragged2e}

\newcommand{\ie}{\emph{i.e., }}
\newcommand{\eg}{\emph{e.g., }}
\newcommand{\etal}{\emph{et al. }}

\newcommand{\wrt}{\emph{w.r.t. }}

\begin{document}
%
\title{Enhancing Stock Movement Prediction with Adversarial Training}
\author{
Fuli Feng\textsuperscript{1}, Huimin Chen\textsuperscript{2}, Xiangnan He\textsuperscript{3}\footnote{Xiangnan He is the corresponding author.}, Ji Ding\textsuperscript{4}, Maosong Sun\textsuperscript{2}, Tat-Seng Chua\textsuperscript{1}\\
$^{1}$National University of Singapore, $^{2}$Tsinghua Unversity,
$^{3}$University of Science and Technology of China, 
$^{4}$University of Illinois at Urbana-Champaign, jiding2@illinois.edu, sms@tsinghua.edu.cn, \{fulifeng93,huimchen1994,xiangnanhe,chuats@gmail.com\}@gmail.com
}
\maketitle
\begin{abstract}


This paper contributes a new machine learning solution for stock movement prediction, which aims to predict whether the price of a stock will be up or down in the near future. The key novelty is that we propose to employ adversarial training to improve the generalization of a neural network prediction model. The rationality of adversarial training here is that the input features to stock prediction are typically based on stock price, which is essentially a stochastic variable and continuously changed with time by nature. As such, normal training with static price-based features (\eg the close price) can easily overfit the data, being insufficient to obtain reliable models. To address this problem, we propose to add perturbations to simulate the stochasticity of price variable, and train the model to work well under small yet intentional perturbations. Extensive experiments on two real-world stock data show that our method outperforms the state-of-the-art solution \cite{xu2018stock} 
with 3.11\% relative improvements on average \wrt accuracy, 
validating the usefulness of adversarial training for stock prediction task.
\end{abstract}
\section{Introduction}
\label{sec:int}

Stock market is one of the largest financial markets, having reached a total value of 80 trillion dollars\footnote{\url{https://data.worldbank.org/indicator/CM.MKT.TRAD.CD?view=chart}.}. Predicting the future status of a stock has always been of great interest to many players in a stock market. While the exact price of a stock is known to be unpredictable~\cite{Walczak:2001,Nguyen:2015}, research efforts have been focused on predicting the stock price movement --- e.g., whether the price will go up/down, or the price change will exceed a threshold --- which is more achievable than stock price prediction~\cite{adebiyi2014comparison,feng2018temporal,xu2018stock}. 

\begin{figure}[tb]
	\subfigure[Training]{
		\label{fig:noshuffle_obj}
		\includegraphics[width=0.23\textwidth]{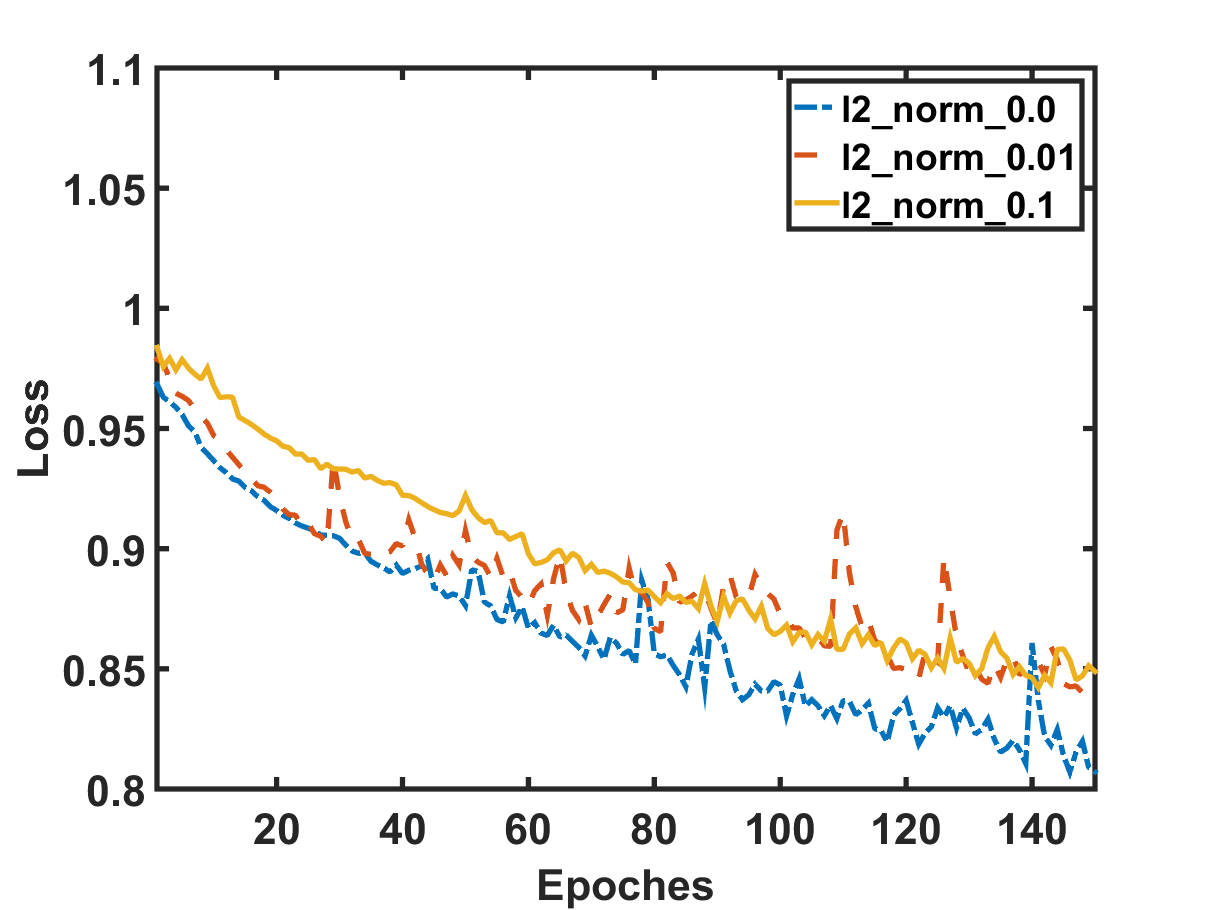}
	}
	\hspace{-0.2in}
	\subfigure[Validation]{
		\label{fig:noshuffle_val}
		\includegraphics[width=0.23\textwidth]{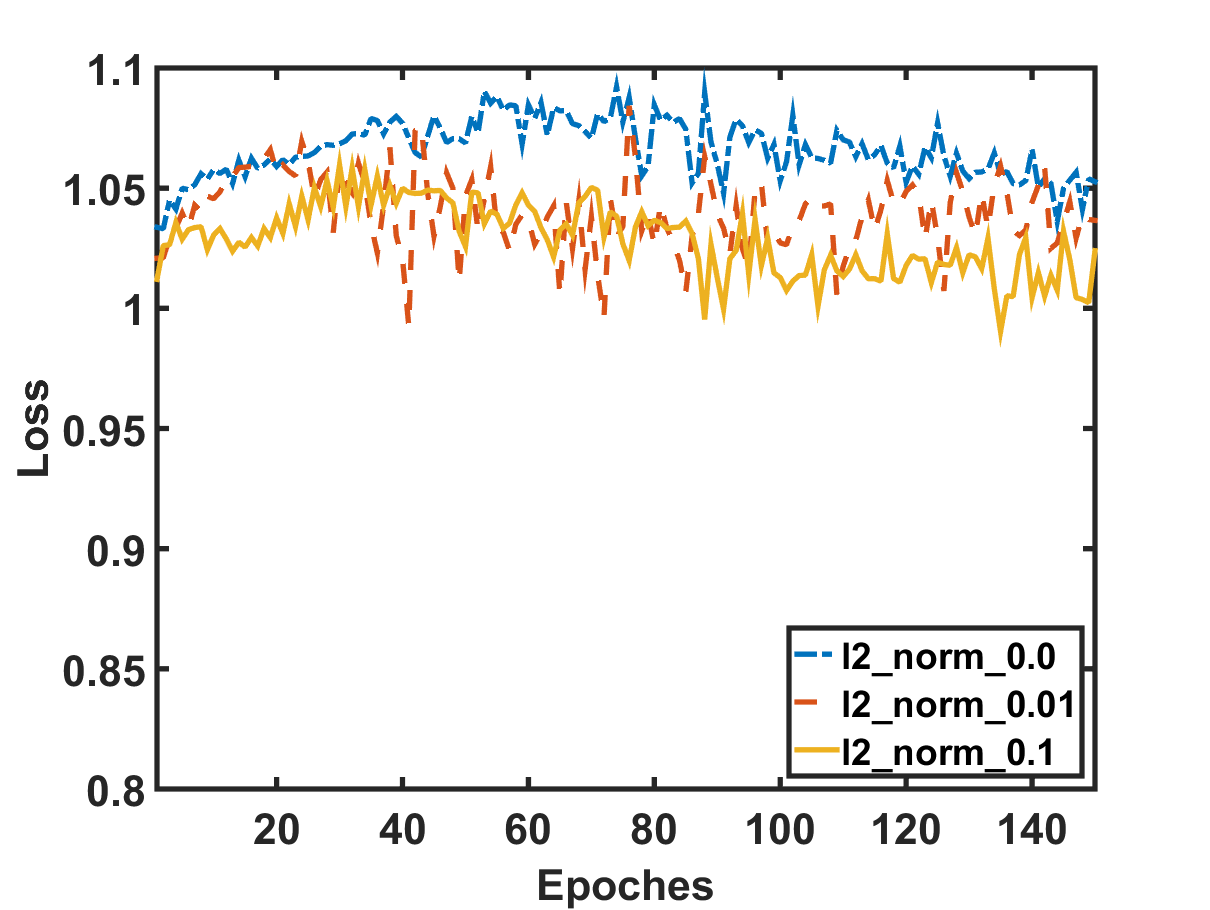}
	}
	\vspace{-0.5cm}
	\caption{Training process of Attentive LSTM with $L_2$ regularization coefficient of $0, 0.01,$ and $0.1$.}
	\label{fig:alstm_gen}
	\vspace{-0.4cm}
\end{figure}
Stock movement prediction can be addressed as a classification task. After defining the label space and features to describe a stock at a time, we can apply standard supervised learning methods such as support vector machines~\cite{huang2005forecasting} and neural networks~\cite{xu2018stock} to build the predictive model. Although technically feasible, we argue that such methods could suffer from weak generalization due to the highly stochastic property of stock market. Figure~\ref{fig:alstm_gen} provides an empirical evidence on the weak generalization, where we split the data into training and validation by time, and train an Attentive LSTM model~\cite{qin2017dual} on the historical prices of stocks to predict their movements. From Figure~\ref{fig:noshuffle_obj}, we can see the training loss gradually decreases with more training epochs, which is as expected. However, the validation loss shown in Figure~\ref{fig:noshuffle_val} does not exhibit a decreasing trend; instead, it only fluctuates around the initialization state without a clear pattern. In other words, the benefits of the model learned on training examples do not translate to improvements on predicting unknown validation examples. We have thoroughly explored the $L_2$ regularization (results of different lines), a common technique to improve model generalization, however, the situation has not improved. 

We postulate the reason is that standard classification methods are assumed to learn from static inputs, such as pixel values in images and term frequencies in documents. When dealing with stochastic variable such as stock price, the static input assumption does not hold and such methods fail to generalize well. Specifically, existing methods for stock prediction typically feed into price-based features, such as the price at a particular time-step or average price on multiple time-steps~\cite{edwards2007technical,nelson2017stock}. Since a stock's price continuously changes with time (during market hours), price-based features are essentially stochastic variables, being fundamentally different from the traditional static inputs. 
To be more specific, the features of a training instance can be seen as a ``sample'' drawn from the distribution of input variables at a particular time-step. Without properly handling the stochasticity of input variables, the method can easily overfit the training data and suffer from weak generalization ability.

In this work, we propose to employ adversarial training to account for the stochastic property of stock market to learn stock movement prediction model. 
Our primary consideration is that given a training example at a particular time-step with fixed input features, the trained model is expected to generate the same prediction on other samples drawn from the inherent distribution of input variables. 
To implement this idea, we can generate additional samples (simulation of the stochasticity) by adding small perturbations on input features, and train the model to perform well on both clean examples and perturbed examples. 
It is the adversarial training method that has been commonly used in computer vision tasks~\cite{kurakin2016adversarial}. However, the problem is that the features to stock prediction models are usually sequential (see Figure 2), such that adding perturbations on the features of all time units can be very time-consuming; moreover, it may cause unintentional interactions among the perturbations of different units which are uncontrollable. To resolve the concern, we instead add perturbations on the high-level prediction features of the model, \eg the last layer which is directly projected to the final prediction. Since most deep learning methods learn abstract representation in the higher layers, their sizes are usually much smaller than the input size. As such, adding perturbations to high-level features is more efficient, and meanwhile it can also retain the stochasticity. 

We implement our adversarial training proposal on an Attentive LSTM model, which is a highly expressive model for sequential data. We add perturbations to the prediction features of the last layer, and dynamically optimize the perturbations to make them change the model's output as much as possible. We then train the model to make it perform well on both clean features and perturbed features. As such, the adversarial training process can be understood as enforcing a dynamic regularizer, which stabilizes the model training and makes the model perform well under stochasticity.

The main contributions of this paper are summarized as: 
\begin{itemize}[leftmargin=*]
	\item We investigate the generalization difficulty in stock movement prediction and highlight the necessity of dealing with the stochastic property of input features. 
	\item We propose an adversarial training solution to address the stochastic challenge, and implement it on a deep learning model for stock movement prediction. 
	\item We conduct extensive experiments on two public benchmarks, validating improvements over several state-of-the-art methods and showing that adversarial learning makes the classifier more robust and more generalizable.
\end{itemize}
\section{Problem Formulation}
\label{sec:met_att}
We use bold capital letters (\eg $\bm{X}$) and bold lower letters (\eg $\bm{x}$) to denote matrices and vectors, respectively. In addition, normal lower case letters (\eg $x$) and Greek letters (\eg $\lambda$) are used to represent scalars and hyper-parameters, respectively. All vectors are in column form, if not otherwise specified. The symbols $tanh$ and $\sigma$ stand for the hyperbolic tangent function and sigmoid function, respectively.

The formulation of stock movement prediction task is to learn a prediction function $\hat{y}^{s} = f(\bm{X}^{s}; \bm{\Theta})$ which maps a stock ($s$) from its sequential features ($\bm{X}^{s}$) to the label space. In other words, the function $f$ with parameters $\bm{\Theta}$ aims to predict the movement of stock $s$ at the next time-step from the sequential features $\bm{X}^{s}$ in the latest $T$ time-steps. $\bm{X}^{s} = [\bm{x}^{s}_{1}, \cdots, \bm{x}^{s}_{T}] \in \mathbb{R}^{D \times T}$ is a matrix which represents the sequential input features (\eg open and close prices, as detailed in Table \ref{tab:mov_fea}) in the lag of past $T$ time-steps, where $D$ is the dimension of features.

Assuming that we have $S$ stocks, we learn the prediction function by fitting their ground truth labels $\mathbf{y} = [y^1, \cdots, y^S] \in \mathbb{R}^{S}$, where $y^s$ (1/-1) is the ground truth label of stock $s$ in the next time-step. We then formally define the problem as:\\
\textbf{Input:} A set of training examples $\{(\bm{X}^s, y^s)\}$.\\
\textbf{Output:} A prediction function $f(\bm{X}^{s}; \bm{\Theta})$, predicting the movement of stock $s$ in the following time-step.

In the practical scenario, we could typically access a long history of each stock, and construct many training examples for each stock by moving the lag along the history. Nevertheless, we use a simplified formulation without loss of generality by only considering one specific lag (\ie one training example for each stock) for briefness of presenting the proposed method.

\section{Adversarial Attentive LSTM (Adv-ALSTM)}
\label{sec:method}
\subsection{Attentive LSTM}
\begin{figure}[]
	\centering
	\includegraphics[width=0.4\textwidth]{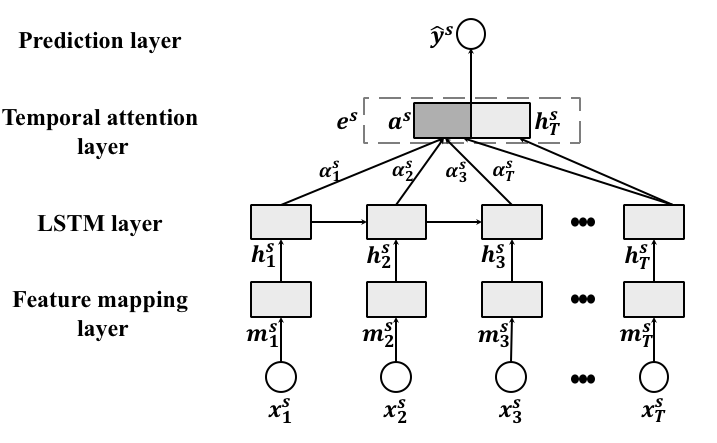}
	\vspace{-0.5cm}
	\caption{Illustration of the \textit{Attentive LSTM}.}
	\vspace{-5pt}
	\label{fig:framework}
\end{figure}

The Attentive LSTM (ALSTM) mainly contains four components: feature mapping layer, LSTM layer, temporal attention, and prediction layer, as shown in Figure \ref{fig:framework}.

\textbf{Feature mapping layer}. Previous work shows that a deeper input gate would benefit the modeling of temporal structures of LSTM \cite{graves2013speech,wu2018parl}. Inspired by their success,  we employ a fully connected layer to project the input features into a \textit{latent representation}. At each time-step, it performs as $\bm{m_t^s} = tanh(\bm{W_m} \bm{x_t^s} + \bm{b_m})$,
which projects the input features to a latent space with dimensionality of $E$. $\bm{W_m} \in \mathbb{R}^{E \times D}$ and $\bm{b_m} \in \mathbb{R}^{E}$ are parameters to be learned. 

\textbf{LSTM layer}. Owing to its ability to capture long-term dependency, LSTM has been widely used to process sequential data~\cite{qin2017dual,chen2018deep}. 
The general idea of LSTM is to recurrently project the input sequence into a sequence of \textit{hidden representations}. At each time-step, the LSTM learns the hidden representation ($\bm{h}^s_t$) by jointly considering the input ($\bm{m}^s_t$) and previous hidden representation ($\bm{h}^s_{t - 1}$) to capture sequential dependency. We formulate it as $\bm{h}^s_t = LSTM(\bm{m}^s_t, \bm{h}^s_{t - 1})$ 
of which the detailed formulation can be referred to \cite{hochreiter1997long}. 
To capture the sequential dependencies and temporal patterns in the historical stock features, an LSTM layer is applied to map $[\bm{m}^{s}_{1}, \cdots, \bm{m}^{s}_{T}]$ into hidden representations $[\bm{h}^{s}_{1}, \cdots, \bm{h}^{s}_{T}] \in \mathbb{R}^{U \times T}$ with the dimension of $U$.

\textbf{Temporal Attention Layer}. The attention mechanism has been widely used in LSTM-based solutions for sequential learning problems～\cite{cho2014learning,chen2018deep}. The idea of attention is to compress the hidden representations at different time-steps into an \textit{overall representation} with adaptive weights. The attention mechanism aims to model the fact that data at different time-steps could contribute differently to the representation of the whole sequence. For stock representation, status at different time-steps might also contribute differently. For instance, days with maximum and minimum prices in the lag might have higher contributions to the overall representation. As such, we use an attention mechanism to aggregate the hidden representations as,
\begin{equation}
\begin{aligned}
\small
	 & \bm{a}^s = \displaystyle{\sum \limits_{t = 1}^{T}} \alpha^s_{t} \bm{h}^s_{t}, ~~ \alpha^s_{t} = \frac{exp^{\widetilde{\alpha}^s_{t}}}{\sum_{t = 1}^{T} exp^{\widetilde{\alpha}^s_{t}}}, \\
	 & \widetilde{\alpha}^s_{t} = \bm{u}_a^T tanh(\bm{W}_a \bm{h}^s_{t} + \bm{b}_a),
\end{aligned}	
\end{equation}
where $\bm{W}_a \in \mathbb{R}^{E' \times U}$, $\bm{b}_a$ and $\bm{u}_a \in \mathbb{R}^{E'}$ are parameters to be learned; and $\bm{a}^s$ is the aggregated representation that encodes the overall patterns in the sequence.

\textbf{Prediction Layer}. Instead of directly making prediction from $\bm{a}^s$, we first concatenate $\bm{a}^s$ with the last hidden state $\bm{h}_T^s$ into the \textit{final latent representation} of stock $s$, 
\begin{align}
    \bm{e}^s = [{\bm{a}^s}^T, {\bm{h}_T^s}^T]^T,
    \label{eq:final_rep}
\end{align}
where $\bm{e}^s \in \mathbb{R}^{2U}$. The intuition behind is to further emphasize the most recent time-step, which is believed to be informative for the following movement \cite{fama2012size}. With $\bm{e}^s$, we use a fully connected layer as the predictive function to estimate the classification confidence $\hat{y}^s = \bm{w}_p^T \bm{e}^s + b_p$. Note that the final prediction is $sign(\hat{y}^s)$.
\subsection{Adversarial Training}
\label{ssec:adv_train}
As with most classification solutions, the \textit{normal} way of training the ALSTM is to minimize an objective function $\Gamma$:
\begin{small}
\begin{align}
	&\sum_{s = 1}^{S}l(y^s, \hat{y}^s) + \frac{\alpha}{2} \|\bm{\Theta}\|_F^2, ~ l(y^s, \hat{y}^s) = max(0, 1 - y^s \hat{y}^s).
	\label{eq:normal_train}
\end{align}
\end{small}
The first term is hinge loss \cite{rosasco2004loss}, which is widely used for optimizing classification models (more reasons of choosing it is further explained in the end of the section). The second term is a regularizer on the trainable parameters to prevent overfitting.

Despite the wide usage of \textit{normal training}, we argue that it is inappropriate for learning stock prediction models. This is because normal training assumes that the inputs are static, ignoring the stochastic property of these features (a training example is a sample drawn from the stochastic distribution of input variables). Note that the features are calculated from stock price, which continuously changes with time and is affected by stochastic trading behaviours at a particular time-step \cite{musgrave1997random}. As such, normal training might lead to model that overfits the data and lacks generalization ability (as shown in Figure \ref{fig:alstm_gen}). Note that is a model performs well under stochasticity would make same predictions for samples drawn from the inherent distribution. Considering that stock price is continuous, our intuition is to intentionally simulate samples by adding small perturbations on static input features. By enforcing the predictions on the simulated samples to be same, the model could capture stochasticity.

\begin{figure}
    \centering
	\includegraphics[width=0.4\textwidth]{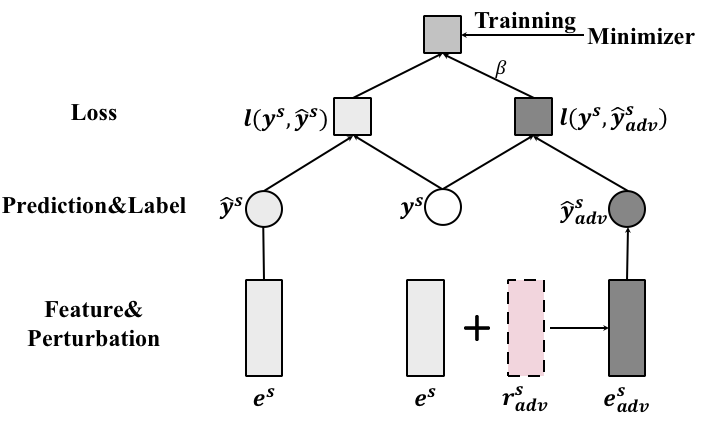}
	\vspace{-0.4cm}
    \caption{Illustration of the \textit{Adversarial Attentive LSTM}.}
    \label{fig:framework_adv}
    \vspace{-0.1cm}
\end{figure}

\textit{Adversarial training} \cite{goodfellow2015explaining,kurakin2016adversarial} implements the aforementioned intuition. It trains a model with both clean examples (\ie examples in the training set) and adversarial examples (AEs) \cite{szegedy2013intriguing}. The AEs are malicious inputs generated by adding intentional perturbations to features of clean examples. The perturbation, named as \textit{adversarial perturbation} (AP) is the direction that leads to the largest change of model prediction. Despite its success in image classification \cite{kurakin2016adversarial}, it is infeasible to be directly applied to stock prediction. This is because calculating perturbations relies on calculation of the gradients regarding the input, which would be time-consuming (caused by the back-propagation through time-step of the LSTM layer). 
Besides, considering the fact that the gradients of the input are dependent across different time-steps, there might be unintentional interactions among the perturbations on different time-steps, which are uncontrollable. To address these problems, we propose to generate AEs from latent representation $\bm{e}^s$, as shown in Figure \ref{fig:framework_adv}.

Before introducing the calculation of AEs, we first elaborate the objective function of Adv-ALSTM:
\begin{align}
\small
	\Gamma_{adv} = \sum_{s = 1}^{S}l(y^s, \hat{y}^s) + \beta \sum_{s = 1}^{S}l(y^s, \hat{y}^s_{adv}) + \frac{\alpha}{2} \|\bm{\Theta}\|_F^2.
	\label{eq:adv_att_lstm}
\end{align}
The second term is an adversarial loss where $\hat{y}^s_{adv}$ is the classification confidence of the AE of stock $s$. $\beta$ is a hyper-parameter to balance the losses of clean and adversarial examples. By minimizing the objective function, the model is encouraged to correctly classify both clean and adversarial examples. Note that a model correctly classifying an AE can make right predictions for examples with arbitrary perturbations at the same scale. This is because AP is the direction leading to the largest change of model prediction. Therefore, adversarial learning could enable ALSTM to capture the stochastic property of stock inputs.

At each iteration, the latent representation of an AE ($\bm{e}^s_{adv}$) is generated by the following formulation,
\begin{align}
\small
	\bm{e}^s_{adv} = \bm{e}^{s} + \bm{r}^{s}_{adv} \text{, } \bm{r}^{s}_{adv} = \arg \max_{\bm{r}^{s}, \| \bm{r}^{s} \| \leq \epsilon} l(y^s, \hat{y}^s_{adv}),
	\label{eq:gen_adv_sample}
\end{align}
where $\bm{e}^s$ (introduced in Equation \ref{eq:final_rep}) is the final latent representation of stock $s$. $\bm{r}^s_{adv}$ is the associated AP. $\epsilon$ is a hyper-parameter to explicitly control the \textit{scale} of perturbation. 
Since it is intractable to directly calculate $\bm{r}^s_{adv}$, we employ the fast gradient approximation method \cite{goodfellow2015explaining}, $\bm{r}^s_{adv} = \epsilon \frac{\bm{g}^s}{\| \bm{g}^s \|} \text{, } \bm{g}^s = \frac{\partial l(y^s, \hat{y}^s)}{\partial \bm{e}^s}.$ 
Specifically, the calculated perturbation is the gradient of loss function regarding the latent representation $\bm{e}^s$ under a $L_2$-norm constraint. Note that the gradient denotes the direction where the loss function increase the most at the given point $\bm{e}^s$, \ie, it would lead to the largest change on the model prediction. 

\begin{figure}[t]
	\centering
	\includegraphics[width=0.32\textwidth]{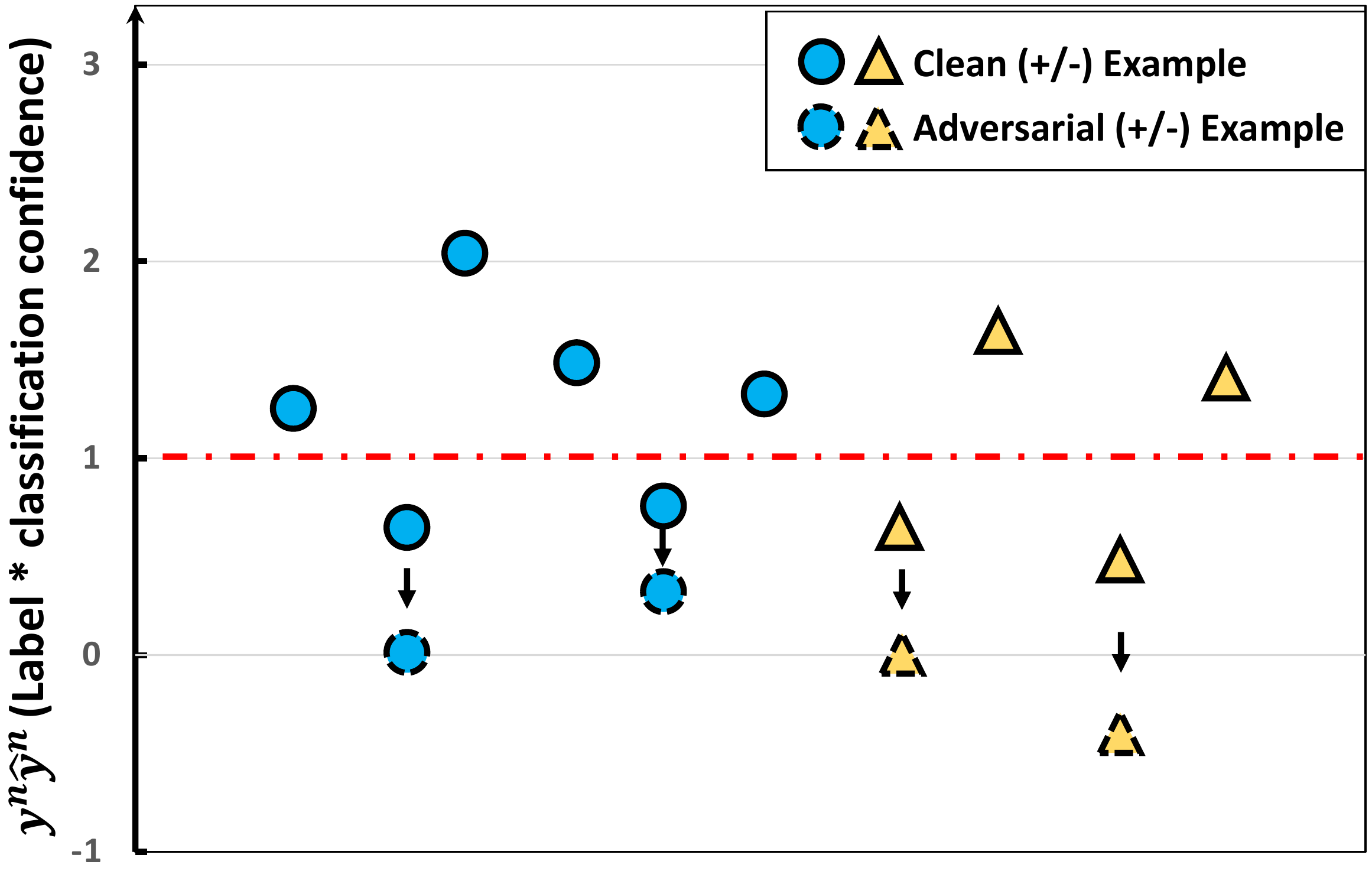} 
	\vspace{-0.4cm}
	\caption{Intuitive illustration of adversarial examples.}
	\vspace{-3pt}
	\label{fig:adv_effect_margin}
\end{figure}

Figure \ref{fig:adv_effect_margin} illustrates the generation of adversarial examples. In a training iteration, given a clean example having loss larger than 0 (\ie $y^s\hat{y}^s < 1$), an AE is generated. The model is then updated to jointly minimize the losses for clean and adversarial examples, which would enforce the margin between clean examples and the decision boundary\footnote{
Minimizing the hinge loss of the AE is adjusting $\bm{w}_p$ to enlarge $y^s \hat{y}^s_{adv} = y^s (\bm{w}_p^T \bm{e}^s + b) + y^s \bm{w}_p^T \bm{r}^s_{adv}$, which would increase the first term $y^s (\bm{w}_p^T \bm{e}^s + b) = y^s \hat{y}^s$. The results in Figure \ref{fig:adv_margin} (in Section \ref{sec:tre}) empirically demonstrate the effect of enforcing margins.}. As such, it would benefit the model to predict examples with perturbations into the same class as the clean one. That is, the model could correctly predict samples drawn from the inherent stochastic distribution of inputs, capturing the stachasticity. While traditional models like support vector machines also push the decision boundary far from clean examples, the adversarial training adaptively adjusts the strength of enforcing margins during the training process since the AP ($\bm{r}^s_{adv}$) varies across iterations. Note that we select the hinge loss to encourage the training process to focus more on the examples close to the decision boundary.

\section{Experiments}
\label{sec:tre}

\subsection{Experimental Settings}
\textbf{Datasets.} We evaluate the proposed method on two benchmarks on stock movement prediction, \textbf{ACL18} \cite{xu2018stock} and \textbf{KDD17} \cite{zhang2017stock}.

\textbf{ACL18} contains historical data from Jan-01-2014 to Jan-01-2016 of 88 high-trade-volume-stocks in NASDAQ and NYSE markets. Following \cite{xu2018stock}, we first align the trading days in the history, \ie removing weekends and public holidays that lack historical prices. We then move a lag with length of $T$ along the aligned trading days to construct candidate examples (\ie one example for a stock on every trading day). We label the candidate examples according to the movement percent of stock close prices\footnote{Given a candidate example of stock $s$ in the lag of $[T' - T + 1, T']$, the movement percent is calculated as $p^s_{T'+1} / p^s_{T'} - 1$, where $p^s_{T'}$ is the adjusted close price of stock $s$ on day $T'$.}. Examples with movement percent $\geq 0.55\%$ and $\leq -0.5\%$ are identified as positive and negative examples, respectively. We temporally split the identified examples into training (Jan-01-2014 to Aug-01-2015), validation (Aug-01-2015 to Oct-01-2015), and testing (Oct-01-2015 to Jan-01-2016).

\textbf{KDD17} includes longer history ranging from Jan-01-2007 to Jan-01-2016 of 50 stocks in U.S. markets. As the dataset is originally collected for predicting stock prices rather than movements, we follow the same approach as \textbf{ACL18} to identify positive and negative examples. We then temporally split the examples into training (Jan-01-2007 to Jan-01-2015), validation (Jan-01-2015 to Jan-01-2016) and testing (Jan-01-2016 to Jan-01-2017).

\begin{table}[!t]
    \caption{Features to describe the daily trend of a stock.}
	\label{tab:mov_fea}
	\resizebox{0.47\textwidth}{!}{%
        \begin{tabular}{c|l}
        \hline
        Features & Calculation \\ \hline\hline
        \textit{c\_open, c\_high, c\_low} & \eg \textit{c\_open} $ = open_t  / close_t - 1$ \\ 
        \textit{n\_close, n\_adj\_close} & \eg \textit{n\_close} = $(close_{t} / close_{t - 1} - 1$ \\ 
        \textit{\begin{tabular}[c]{@{}c@{}}5-day, 10-day, 15-day,\\ 20-day, 25-day, 30-day\end{tabular}} & \eg \textit{5-day} = $\frac{\sum_{i=0}^4 adj\_close_{t - i} / 5}{adj\_close_{t}} - 1$ \\ \hline
\end{tabular}%
}
\end{table}

\textbf{Features.} Instead of using the raw EOD data, we define 11 temporal features ($\mathbf{x}_t^s$) to describe the trend of a stock $s$ at trading day $t$. Table \ref{tab:mov_fea} elaborates the features associated with calculation. Our aim of defining these features are to: 1) normalize the prices of different stocks; 2) and explicitly capture the interaction of different prices (\eg open and close). 
%

\textbf{Baselines.} We compare the following methods:
\begin{itemize}[leftmargin=*]
	\item \textbf{MOM} Momentum (MOM) is a technical indicator that predicts negative or positive for each example with the trend in the last 10 days.
	\item \textbf{MR} Mean reversion (MR) predicts the movement of each example as the opposite direction of latest price towards the 30-day moving average.
	\item \textbf{LSTM} is a neural network with an LSTM layer and a prediction layer \cite{nelson2017stock}. We tune three hyper-parameters, number of hidden units ($U$), lag size ($T$), and weight of regularization term ($\lambda$).
	\item \textbf{ALSTM} is the \textit{Attentive LSTM}~\cite{qin2017dual}, which is optimized with normal training. Similar as \textbf{LSTM}, we also tune $U$, $T$, and $\lambda$.
	\item \textbf{StockNet} uses a Variational Autoencoder (VAE) to encode the stock input so as to capture the stochasticity, and a temporal attention to model the importance of different time-steps \cite{xu2018stock}. Here we take our temporal features in Table \ref{tab:mov_fea} as inputs and tune its hidden size, dropout ratio, and auxiliary rate ($\alpha$).
\end{itemize}

\textbf{Evaluation Metrics.} We evaluate the prediction performance with two metrics, Accuracy (Acc) and Matthews Correlation Coefficient (MCC)~\cite{xu2018stock} of which the ranges are in $[0, 100]$ and $[-1, 1]$. Note that better performance is evidenced by higher value of the metrics.

\textbf{Parameter Settings.} We implement the \textbf{Adv-ALSTM} with Tensorflow and optimize it using the mini-batch Adam\cite{Kingma2015Adam} with a batch size of 1,024 and an initial learning rate of 0.01. We search the optimal hyper-parameters of \textbf{Adv-ALSTM} on the validation set. For $U$, $T$, and $\lambda$, \textbf{Adv-ALSTM} inherits the optimal settings from \textbf{ALSTM}, which are selected via grid-search within the ranges of [4, 8, 16, 32], [2, 3, 4, 5, 10, 15], and [0.001, 0.01, 0.1, 1], respectively. We further tune $\beta$ and $\epsilon$ within [0.001, 0.005, 0.01, 0.05, 0.1, 0.5, 1] and [0.001, 0.005, 0.01, 0.05, 0.1], respectively. We report the mean testing performance when \textbf{Adv-ALSTM} performs best on the validation set over five different runs. 
Code could be accessed through \url{https://github.com/hennande/Adv-ALSTM}.

\subsection{Experimental Results}
\textbf{Performance Comparison.} Tables \ref{tab:perf_acl18} shows the prediction performance of compared methods on the two datasets regarding Acc and MCC, respectively. From the table, we have the following observations:
\begin{itemize}[leftmargin=*]
	\item \textbf{Adv-ALSTM} achieves the best results in all the cases. Compared to the baselines, \textbf{Adv-ALSTM} exhibits an improvement of 4.02\% and 42.19\% (2.14\% and 56.12\%) on the \textbf{ACL18} (\textbf{KDD17}) dataset regarding Acc and MCC, respectively. This justifies the effectiveness of adversarial training, which might be due to enhancing the model generalization via adaptively simulating perturbations during the training. 
	\item Specifically, compared to \textbf{StockNet}, which captures stochasticity of stock inputs with VAE, \textbf{Adv-ALSTM} achieves significant improvements. We postulate the reason is that \textbf{StockNet} cannot explicitly model the scale and direction of stochastic perturbation since it relies on Monte Carlo sampling during the training process.
	\item Among the baselines, \textbf{ALSTM} outperforms \textbf{LSTM} by 1.93\% and 48.69\% on average \wrt Acc and MCC, which validates the impact of attention~\cite{qin2017dual}. Besides, \textbf{MOM}  and \textbf{MR} performs worse than all the machine learning-based methods as expected, which justifies that historical patterns help in stock prediction task. 
\end{itemize}

\begin{table}[t]
	\centering
	\caption{Performance comparison on the two datasets.}
	\label{tab:perf_acl18}
	\resizebox{0.47\textwidth}{!}{%
		\begin{tabular}{c||cc|cc}
			\hline
			\multirow{2}{*}{Method} & \multicolumn{2}{c|}{ACL18} & \multicolumn{2}{c}{KDD17} \\ 
 & Acc & MCC & Acc & MCC \\ \hline \hline
			\textbf{MOM} & 47.01$\pm$----- & -0.0640$\pm$----- & 49.75$\pm$----- & -0.0129$\pm$-----\\ 
			\textbf{MR} & 46.21$\pm$----- & -0.0782$\pm$----- & 48.46$\pm$----- & -0.0366$\pm$----- \\ 
			\textbf{LSTM} & 53.18$\pm$5e-1 & 0.0674$\pm$5e-3 & 51.62$\pm$4e-1 & 0.0183$\pm$6e-3 \\ 
			\textbf{ALSTM} & 54.90$\pm$7e-1 & 0.1043$\pm$7e-3 & 51.94$\pm$7e-1 & 0.0261$\pm$1e-2 \\ 
			\textbf{StockNet} & 54.96$\pm$----- & 0.0165$\pm$----- & 51.93$\pm$4e-1 & 0.0335$\pm$5e-3 \\ 
			\textbf{Adv-ALSTM} & \textbf{57.20$\pm$-----} & \textbf{0.1483$\pm$-----} & \textbf{53.05$\pm$-----} & \textbf{0.0523$\pm$-----} \\ \hline
			RI & 4.02\% & 42.19\% & 2.14\% & 56.12\% \\ \hline
		\end{tabular}%
	}
	\justify
	\scriptsize{RI denotes the relative improvement of \textbf{Adv-ALSTM} compared to the best baseline. The performance of \textbf{StockNet} is directly copied from \cite{xu2018stock}.}
\vspace{-0.5cm}
\end{table}

\begin{table}[!t]
	\centering
	\caption{Performance of \textbf{Rand-ALSTM} on the two datasets.}
	\label{tab:perf_rand}
	\resizebox{0.3\textwidth}{!}{%
	\begin{tabular}{c||c|c}
		\hline
		Datasets & Acc & MCC  \\ \hline \hline
		\textbf{ACL18} & 55.08$\pm$2e0 & 0.1103$\pm$4e-2 \\
		\textbf{KDD17} & 52.43$\pm$5e-1 & 0.0405$\pm$8e-3 \\ \hline
	\end{tabular}%
	}
\vspace{-0.3cm}
\end{table}

\begin{figure}[t]
	\centering
	\mbox{
		\subfigure[Validation of \textbf{ACL18}]{
			\label{fig:acl_margin_val}
			\includegraphics[width=0.23\textwidth]{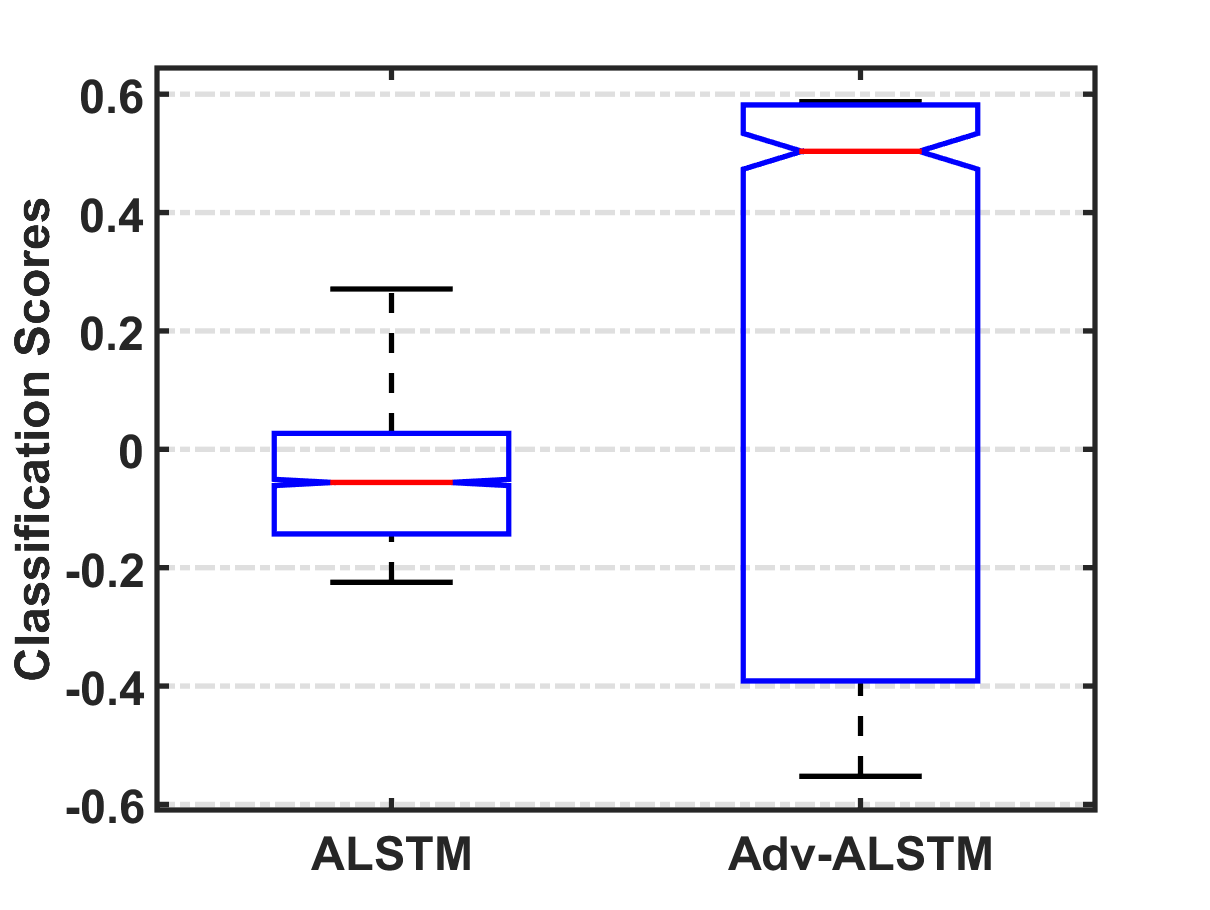}
		}
		\hspace{-0.2in}
		\subfigure[Testing of \textbf{ACL18}]{
			\label{fig:acl_margin_tes}
			\includegraphics[width=0.23\textwidth]{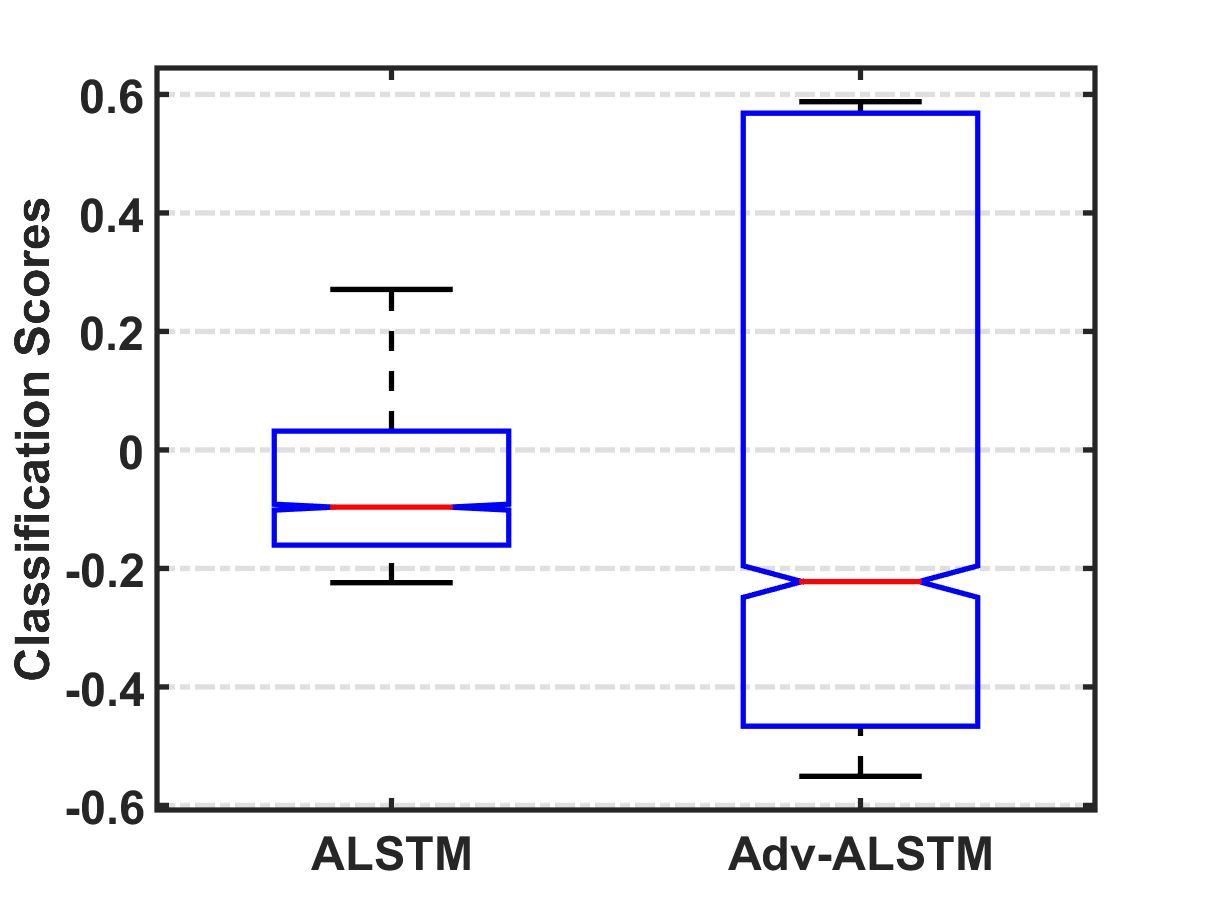}
		}
	}
	\vspace{-0.4cm}
	\caption{Distributions of classification confidences assigned by \textbf{ALSTM} and \textbf{Adv-ALSTM} for clean examples.
	}
	\label{fig:adv_margin}
	\vspace{-0.4cm}
\end{figure}

\begin{figure}[t]
	\centering
	\mbox{
		\subfigure[Acc]{
			\label{fig:acl_adv_rel_dec_acc}
			\includegraphics[width=0.23\textwidth]{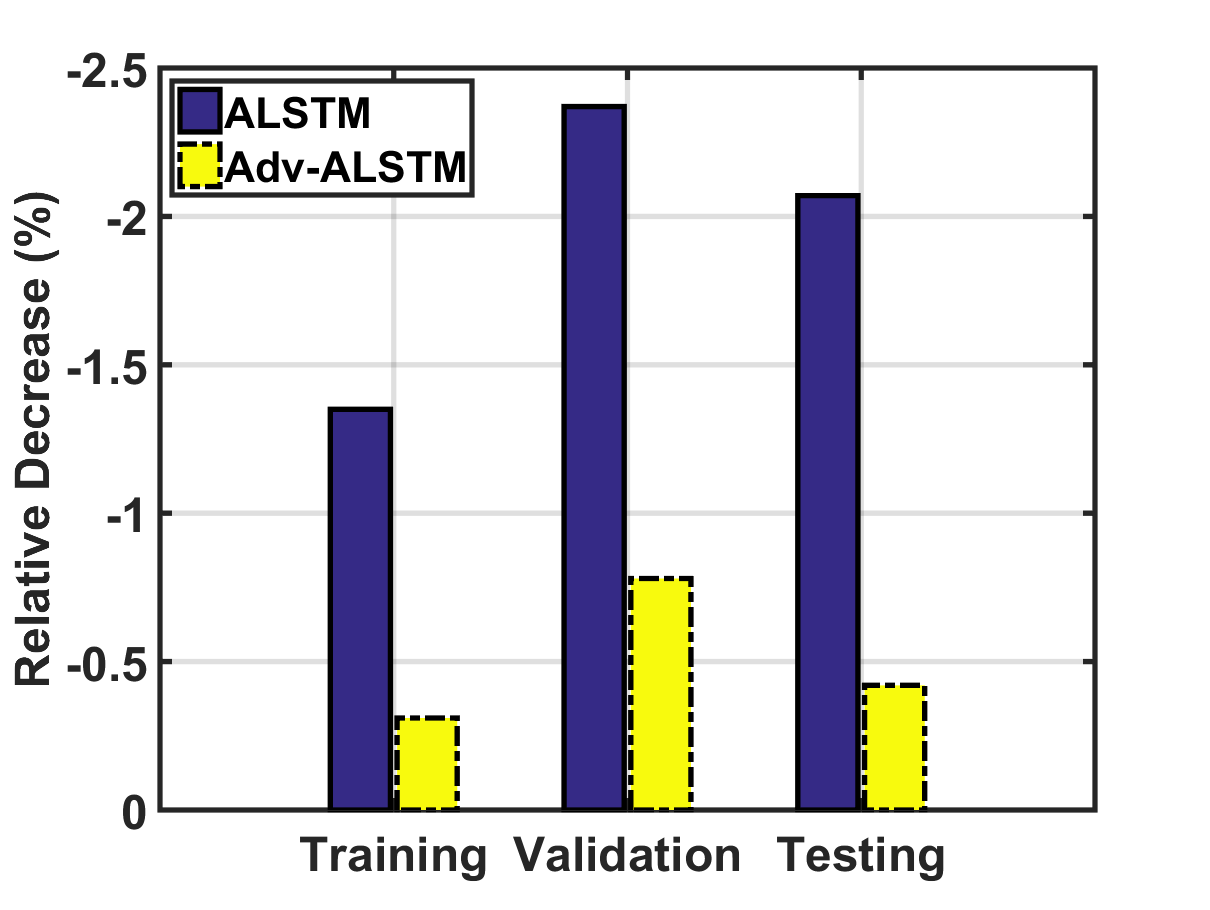}
		}
		\hspace{-0.2in}
		\subfigure[MCC]{
			\label{fig:acl_adv_rel_dec_mcc}
			\includegraphics[width=0.23\textwidth]{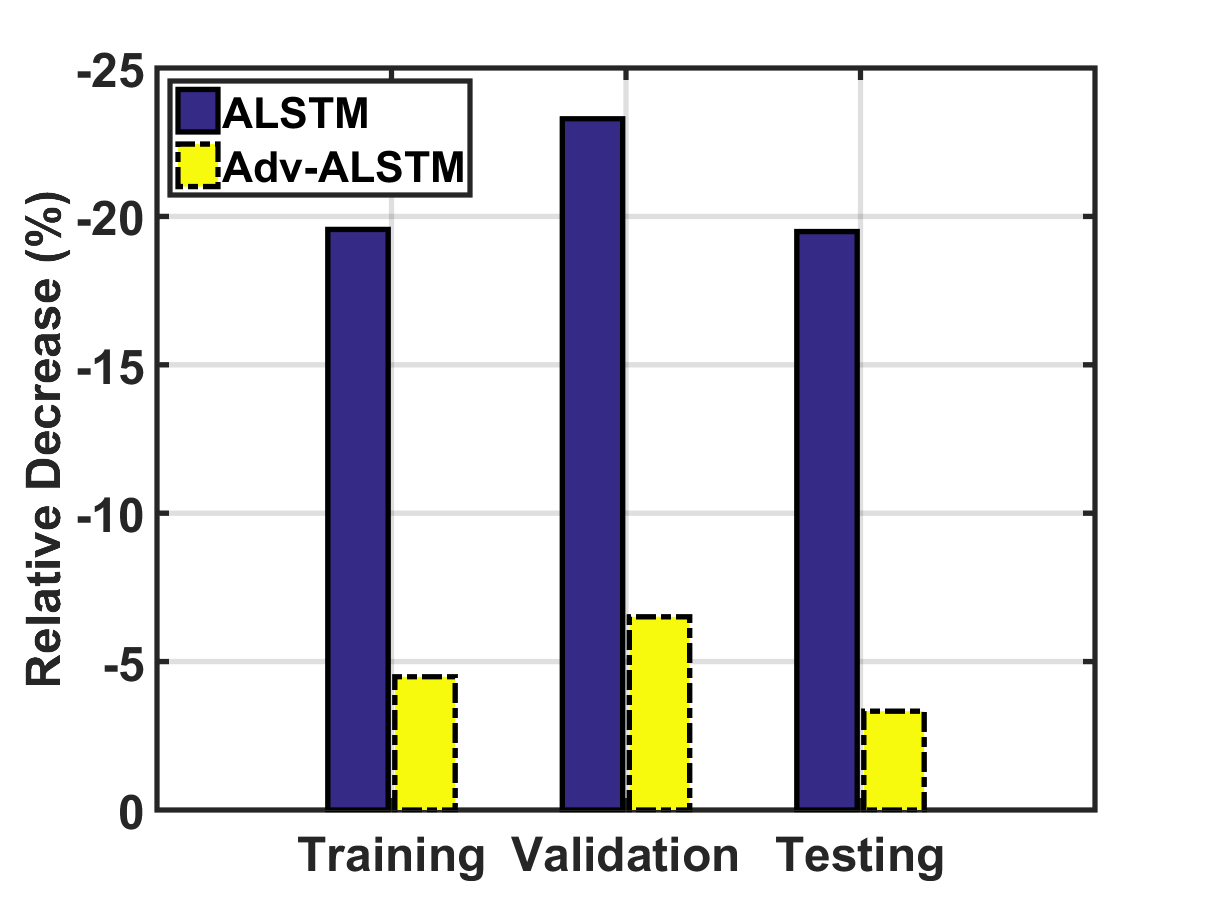}
		}
	}
	\vspace{-0.4cm}
	\caption{Robustness against adversarial example of \textbf{ALSTM} and \textbf{Adv-ALSTM}. Each plotted number is the RPD of a model on adversarial examples compared to clean ones.
	}
	\label{fig:adv_perf_decrease}
	\vspace{-0.3cm}
\end{figure}

\subsubsection{Stochastic Perturbation \textit{VS.} Adversarial Perturbation.}
We further investigate the effectiveness of adversarial training via comparing adversarial perturbations and random ones. \textbf{Rand-ALSTM} is a variance of \textbf{Adv-ALSTM}, which generates additional examples by adding random perturbations to the input of clean examples. Table \ref{tab:perf_rand} shows the performance of \textbf{Rand-ALSTM} on the two datasets. By cross comparing it with Table \ref{tab:perf_acl18}, we observe that: 1) Compared to \textbf{Rand-ALSTM}, \textbf{Adv-ALSTM} achieves significant improvements. For instance, its performance \wrt Acc on \textbf{ACL18} is 3.95\% better than that of \textbf{Rand-ALSTM}. It demonstrates that adversarial perturbations are helpful for stock prediction, similar to that reported in the original image classification tasks \cite{goodfellow2015explaining}. 2) \textbf{Rand-ALSTM} outperforms \textbf{ALSTM}, which is purely trained with clean examples, with an average improvement of 0.64\% \wrt Acc on the two datasets. This highlights the necessity of dealing with stochastic property of stock features.

\textbf{Impacts of Adversarial Training.} We now investigate the impacts of adversarial training to answer: 1) Whether the adversarial training \textit{enforces the margin} between clean examples and the decision boundary. 2) Whether the adversarial training \textit{enhances the robustness} of the model against adversarial examples.
Note that we only show the results on the \textbf{ACL18} dataset as the results on \textbf{KDD17} admit the same observations.

\textit{Enforcing margin.} Recall that the only difference between \textbf{Adv-ALSTM} and \textbf{ALSTM} is learning parameters with adversarial training and standard training. As such, we answer the first question by comparing the classification confidence of clean examples (larger value denotes larger margin to the decision boundary) assigned by \textbf{Adv-ALSTM} and \textbf{ALSTM}. Figure \ref{fig:adv_margin} illustrates the distributions of the classification confidences assigned by \textbf{ALSTM} and \textbf{Adv-ALSTM}. As can be seen, the confidences of \textbf{Adv-ALSTM} distribute in a range ([-0.6, 0.6] roughly), which is about 1.5 times larger than that of \textbf{ALSTM} ([-0.2, 0.3]). It indicates that adversarial training pushes the decision boundary far from clean examples, which is believed to help enhance the robustness and generalization ability of the model.

\textit{Robustness against adversarial examples.} We then investigate the second question via comparing the performance of \textbf{ALSTM} and \textbf{Adv-ALSTM} on the clean and associated adversarial examples. Figures \ref{fig:acl_adv_rel_dec_acc} and \ref{fig:acl_adv_rel_dec_mcc} illustrate the relative performance decrease (RPD) of \textbf{ALSTM} and \textbf{Adv-ALSTM} on adversarial examples regarding the one on clean examples, respectively. Note that larger absolute value of RPD indicates that the model is more vulnerable to adversarial perturbations. As can be seen, the average RPD of \textbf{ALSTM} is 4.31 (6.34) times larger as compared to \textbf{Adv-ALSTM} regarding Acc (MCC). This justifies the potential of enhancing model robustness with adversarial training. 

\section{Related Work}
\label{sec:rel}
\subsection{Stock Movement Prediction}
Recent works on stock movement prediction, mainly fall under two categories, \textit{technical analysis} and \textit{fundamental analysis} (FA). The \textit{technical analysis} (TA) takes historical prices of a stock as features to forecast its movement. 
Most of recent methods in TA mine stock movements with deep models \cite{lin2017hybrid,nelson2017stock,chong2017deep}. Among them, recurrent neural networks like LSTM have become key components to capture the temporal patterns of stock prices \cite{nelson2017stock,lin2017hybrid}. Besides, other advanced neural models, such as convolution neural networks (CNN) \cite{lin2017hybrid} and deep Boltzmann machine \cite{chong2017deep}, are also evidenced to be beneficial for capturing the non-linearity of stock prices.

In addition to price features, FA also examines related economic, financial, and other qualitative and quantitative factors \cite{hu2018listening,zhang2018improving,li2018web,xu2018stock}. For instance, Xu and Cohen \shortcite{xu2018stock} incorporate signals from social media, which reflects opinions from general users, to enhance stock movement prediction. Specifically, they employ a VAE to learn a stock representation by jointly encoding the historical prices and tweets mentioning it. Moreover, Zhang \etal \shortcite{zhang2018improving} further consider news events related to a stock or the associated company via a coupled matrix and tensor factorization framework.

Both TA and FA studies show that price features play crucial roles in stock movement prediction. However, most of the existing works assume stock price as stationary, which thus lack the ability to deal with its stochastic property. StockNet~\cite{xu2018stock} is the only exception which tackles this problem via VAE. VAE encodes the inputs into a latent distribution and enforces samples from the latent distribution to be decoded with the same prediction. Generally, the philosophy behind is similar as the simulation of stochastic perturbations since one sample from the latent distribution can be seen as adding stochastic perturbation to the latent representation. 
As compared to our method, our perturbation is intentionally generated which indicates leads to hardest examples for the model to obtain the target prediction. In addition, the proposed method can be easily adapted to other solutions of stock movement predictions.

\subsection{Adversarial Learning}
Adversarial learning has been intensively studied by training a classification model to defense adversarial examples, which are intentionally generated to perturb the model. Existing works of adversarial learning mainly concentrate on computer vision tasks like image classification \cite{goodfellow2015explaining,miyato2016adversarial,kurakin2016adversarial,yang2018shuffle,chen2018zero}. Owing to the property that image features are typically continued real values, adversarial examples are directly generated in the feature space. Recently, several works extend the adversarial learning to tasks with discrete inputs such as text classification (a sequence of words) \cite{miyato2016adversarial}, recommendation (user and item IDs) \cite{he2018adversarial}, and graph node classification (graph topology) \cite{dai2018adversarial,feng2019graph}. Rather than in the feature space, these works generate adversarial examples from embedding of inputs such as word, user (item), and node embeddings. Although this work is inspired by these adversarial learning research efforts, it targets a distinct task---stock movement prediction, of which the data are time series with stochastic property. To the best of our knowledge, this work is the first one to explore the potential of adversarial training in time-series analytics.
\section{Conclusion}
\label{sec:con}
In this paper, we showed that neural network solutions for stock movement prediction could suffer from weak generalization ability since they lack the ability to deal with the stochasticity of stock features. To solve this problem, we proposed an \textit{Adversarial Attentive LSTM} solution, which leverages adversarial training to simulate the stochasticity during model training. We conducted extensive experiments on two benchmark datasets and validated the effectiveness of the proposed solution, signifying the importance of accounting for the stochasticity of stock prices in stock movement prediction. Morever, the results showed that adversarial training enhances the robustness and generalization of the prediction model.

In future, we plan to explore the following directions: 1) we are interested in testing \textbf{Adv-ALSTM} in movement prediction of more assets such as commodities. 2) We plan to apply adversarial training to stock movement solutions with different structures such as the CNNs \cite{lin2017hybrid}. 3) We will explore the effect of adversarial training over fundamental analysis methods of stock movement prediction.
\bibliographystyle{named}
\bibliography{main}
\end{document}